# Quantum spin Hall effect of light

Konstantin Y. Bliokh[1,2], Daria Smirnova[2], and Franco Nori[1,3]

[1]*Center for Emergent Matter Science, RIKEN, Wako-shi, Saitama 351-0198, Japan*
[2]*Nonlinear Physics Centre, RSPhysE, The Australian National University, ACT 0200, Australia*
[3]*Physics Department, University of Michigan, Ann Arbor, Michigan 48109-1040, USA*

**Maxwell's equations, formulated 150 years ago, ultimately describe properties of light, from classical electromagnetism to quantum and relativistic aspects. The latter ones result in remarkable geometric and topological phenomena related to the spin-1 massless nature of photons. By analyzing fundamental spin properties of Maxwell waves, we show that free-space light exhibits an intrinsic quantum spin Hall effect, i.e., surface modes with strong spin-momentum locking. These modes are evanescent waves that form, e.g., surface plasmon-polaritons at vacuum-metal interfaces. Our findings illuminate the unusual transverse spin in evanescent waves and explain recent experiments demonstrating the transverse spin-direction locking in the excitation of surface optical modes. This deepens our understanding of Maxwell's theory, reveals analogies with topological insulators for electrons, and offers applications for robust spin-directional optical interfaces.**

Solid-state physics exhibits a family of Hall effects with remarkable physical properties. The usual Hall effect (HE) and quantum Hall effect (QHE) appear in the presence of an external magnetic field, which breaks the time-reversal ($\mathcal{T}$) symmetry of the system. The HE induces charge current orthogonal to both the magnetic field and an applied electric field, whereas the QHE [1] involves distinct topological electron states, with unidirectional edge modes (charge-momentum locking), characterized by the topological Chern number [2].

The intrinsic spin Hall effect (SHE) can occur in $\mathcal{T}$-symmetric electron systems with spin-orbit interactions. It produces a spin-dependent transport of electrons orthogonal to the external driving force [3,4]. There is also the quantum spin Hall effect (QSHE) [5,6], which is characterized by unidirectional edge spin transport, i.e., edge states with opposite spins propagating in opposite directions. Such topological states with spin-momentum locking gave rise to a new class of materials: topological insulators [7,8].

Alongside the extensive condensed-matter studies of electron Hall effects, their photonic counterparts have been found in various optical systems. In particular, both the HE [9] the QHE with unidirectional edge propagation [10,11] have been reported in magneto-optical systems with broken $\mathcal{T}$-symmetry. Furthermore, because photons are relativistic spin-1 particles, they naturally exhibit intrinsic spin-orbit interaction effects, including Berry phase [12] and the SHE [13–15] stemming from fundamental spin properties of Maxwell equations [16].

The only missing part in the above optical Hall effects is the QSHE for photons. Recently, it was suggested that photonic topological insulators can be created in complex metamaterials structures [17–19]. Here we show that pure free-space light already possesses intrinsic QSHE, and simple natural materials (such as metals supporting surface plasmon-polariton modes) exhibit some features resembling topological insulators. We show that the recently discovered transverse spin in evanescent waves [20,21] and spin-controlled unidirectional excitation of surface or waveguide modes [22–27] can be interpreted as manifestations of the QSHE of light.

Propagating (bulk) free-space modes of Maxwell equations are polarized plane waves. Introducing the complex amplitude $\mathbf{E}(\mathbf{r})$ of the harmonic electric field $\mathcal{E}(\mathbf{r},t) = \text{Re}\left[\mathbf{E}(\mathbf{r})e^{-i\omega t}\right]$, the plane-wave solution with wave vector $\mathbf{k} = k\bar{\mathbf{z}}$ is

$$\mathbf{E} \propto \mathbf{e}\exp(ikz), \quad \mathbf{e} = \alpha\overline{\mathbf{x}} + \beta\overline{\mathbf{y}}. \tag{1}$$

Here $k = \omega/c$, $\mathbf{e}$ is the complex unit polarization vector ($|\alpha|^2 + |\beta|^2 = 1$), whereas $\overline{\mathbf{x}}$, $\overline{\mathbf{y}}$, and $\overline{\mathbf{z}}$ denote the unit vectors of the corresponding axes. The Jones vector $\xi = (\alpha, \beta)^T$ is a three-dimensional spinor, which describes the SU(2) polarization state of light. The spin states of propagating light are circular polarizations $\xi = (1, \pm i)^T/\sqrt{2}$ with helicities $\sigma \equiv 2\,\text{Im}(\alpha^*\beta) = \pm 1$. According to the massless nature of photons, the plane-wave spin is directed along the wave vector: $\mathbf{S} = \sigma\mathbf{k}/k$ (we consider the spin density per photon in $\hbar = 1$ units; see Supplementary Text).

Generalizing Eq. (1) to an arbitrary direction of propagation, the polarization vector becomes momentum-dependent: $\mathbf{e}(\mathbf{k})$. Namely, it is tangent to the $\mathbf{k}$-space sphere due to the transversality condition $\mathbf{E} \cdot \mathbf{k} = 0$. This spherical $\mathbf{k}$-space geometry underlies the spin-orbit interaction of light [12–16]. In particular, introducing the helicity basis of circular polarizations $\mathbf{e}^\sigma(\mathbf{k})$ [16], one can calculate the Berry connection $\mathbf{A}^{\sigma\sigma'} = -i\mathbf{e}^\sigma \cdot (\nabla_\mathbf{k})\mathbf{e}^{\sigma'}$ and curvature $\mathbf{F}^{\sigma\sigma'} = \nabla_\mathbf{k} \times \mathbf{A}^{\sigma\sigma'}$ for photons. In agreement with the helicity-degenerate light-cone spectrum of photons, the Berry curvature is diagonal, $\mathbf{F}^{\sigma\sigma'} = \delta^{\sigma\sigma'}\mathbf{F}^\sigma$, and it forms two monopoles at the Dirac-point origin of the momentum space [12–16]:

$$\mathbf{F}^\sigma = \sigma\frac{\mathbf{k}}{k^3}, \quad \sigma = \pm 1. \tag{2}$$

This curvature is responsible for the spin-redirection Berry phase and the SHE in optics [12–16].

We now define the topological Chern numbers for the two helicity states: $C^\sigma = \frac{1}{2\pi}\oint \mathbf{F}^\sigma d^2\mathbf{k}$, where the integral is taken over the $\mathbf{k}$-space sphere. The Chern numbers are meaningful in systems with Abelian Berry phases, such as two-dimensional systems with the conserved spin component along the third dimension [7,8,28]. This is also the case for photons having Abelian Berry phase, two-dimensional polarization on the $\mathbf{k}$-space sphere, and conserved radial $\mathbf{k}$-component of the spin (i.e., helicity) [29]. The monopole curvature (2) yields $C^\sigma = 2\sigma$. The total Chern number $C = \sum_{\sigma=\pm 1} C^\sigma$ and the spin Chern number $C_{\text{spin}} = \sum_{\sigma=\pm 1} \sigma C^\sigma$ characterize the photonic QHE and QSHE properties [7,8,28]:

$$C = 0, \quad C_{\text{spin}} = 4. \tag{3}$$

The physical meaning of the Chern numbers is the number of edge modes with fixed direction of propagation. The vanishing total Chern number (3) reflects the $\mathcal{T}$-symmetry of Maxwell equations and the absence of the QHE for free-space photons. At the same time, the non-zero spin Chern number (3) implies that free-space light has two pairs of QSHE modes, i.e., edge counter-propagating modes with opposite spins. Furthermore, the value $C_{\text{spin}} = 4$ implies that the topological $\mathbb{Z}_2$ invariant, associated with the $\mathcal{T}$-symmetry, vanishes: $\nu = \frac{C_{\text{spin}}}{2}\,\text{mod}\,2 = 0$. This means that surface modes of Maxwell equations are not helical fermions [30] as, e.g., surface states of the Dirac equation [31,32].

Nonetheless, nontrivial QSHE states of light exist, and they are well known. The photonic edge states of a bounded segment of free-space are evanescent waves. For instance, assuming the $x = 0$ boundary, with free space at $x > 0$, the generic evanescent-wave solution of Maxwell equations can be written as [21]

$$\mathbf{E}_{evan} \propto \mathbf{e}_{evan} \exp(ik_z z - \kappa x), \quad \mathbf{e}_{evan} = \alpha \overline{\mathbf{x}} + \beta \frac{k}{k_z} \overline{\mathbf{y}} - i\alpha \frac{\kappa}{k_z} \overline{\mathbf{z}}. \qquad (4)$$

Here spinor $\xi = (\alpha, \beta)^T$ still characterizes the wave polarization states. The wave (4) propagates along the $z$-axis with wave number $k_z > k$ and decays exponentially away from the boundary with the decrement $\kappa = \sqrt{k_z^2 - k^2}$.

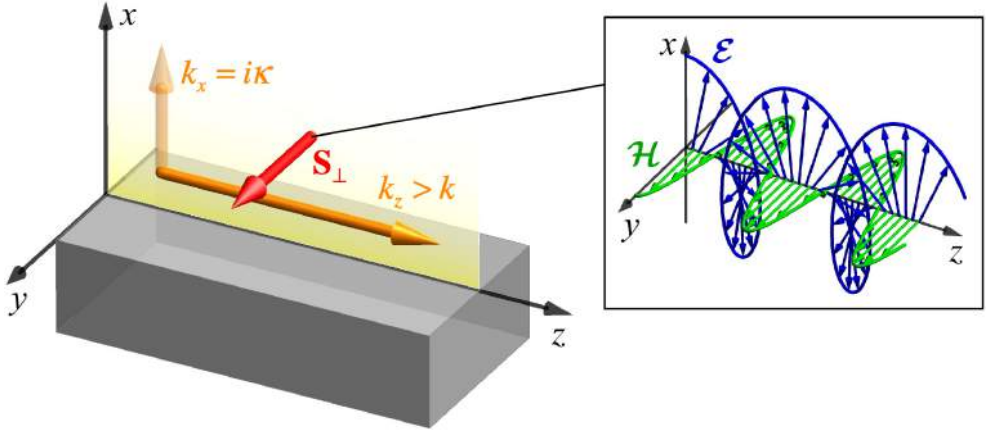

**Fig. 1. Transverse spin in evanescent electromagnetic waves.** The evanescent wave (4) propagates along the $z$-axis and decays exponentially in the $x > 0$ semi-space. The inset shows the instantaneous distributions of the electric and magnetic wave fields for the case of linear transverse-magnetic polarization, $\xi = (1,0)^T$. The cycloidal $(x,z)$-plane rotation of the electric field generates the transverse spin (5) $\mathbf{S}_\perp$ [20,21]. The sign of the transverse spin depends on the direction of propagation of the evanescent wave.

One can consider the evanescent wave (4) as a plane wave with the complex wave vector $\mathbf{k} = k_z \overline{\mathbf{z}} + i\kappa \overline{\mathbf{x}}$. Importantly, the transversality condition $\mathbf{E} \cdot \mathbf{k} = 0$ generates the imaginary longitudinal $z$-component in the polarization vector $\mathbf{e}_{evan}$, in contrast to the purely transverse polarization $\mathbf{e}$ in propagating waves (1). This component produces $(x,z)$-plane rotation of the electric or magnetic fields and thereby generates unusual transverse spin in evanescent waves [20,21], Fig. 1. This transverse spin is independent of the polarization $\xi$ and can be written as

$$\mathbf{S}_\perp = \frac{\operatorname{Re}\mathbf{k} \times \operatorname{Im}\mathbf{k}}{(\operatorname{Re}\mathbf{k})^2}, \qquad (5)$$

Remarkably, Eq. (5) demonstrates spin-momentum locking, similar to that in QSHE and 3D topological insulators for electrons [5–8]. In particular, the $z$-propagating evanescent waves with $k_z > 0$ and $k_z < 0$ will have opposite transverse spins $S_y > 0$ and $S_y < 0$ (Figs. 2–4). Thus, any interface between free-space and a medium supporting surface or guided modes with evanescent tails (4) exhibits counter-propagating opposite-spin edge modes, i.e., the QSHE of light. This is the first key point of our work.

In agreement with $C_{spin} = 4$, there are two pairs of the QSHE modes in free space, because the evanescent waves (4) are double-degenerate with respect to helicities $\sigma = \pm 1$. However, the existence of surface modes in Maxwell equations requires a planar interface between vacuum and a medium characterized by a permittivity $\varepsilon$ and permeability $\mu$. Such interface breaks the dual symmetry between the electric and magnetic properties: $\varepsilon \neq \mu$ [29]. This breaks the polarization degeneracy, and only a single polarization survives in surface modes. For example, only transverse-magnetic surface waves exist at the interface with a medium with $\mu = 1$ and $\varepsilon < -1$. Calculating the spectrum, polarization, and spin of these surface modes of Maxwell equations, we obtain (see Supplementary Text):

$$\omega_{\text{surf}} = \sqrt{\frac{1+\varepsilon}{\varepsilon}} k_{\text{surf}}, \quad \xi_{\text{surf}} = \begin{pmatrix} 1 \\ 0 \end{pmatrix}, \quad \langle \mathbf{S}_{\text{surf}} \rangle = \frac{1}{\sqrt{-\varepsilon}} \bar{\mathbf{k}}_{\text{surf}} \times \bar{\mathbf{n}}. \quad (6)$$

Here $\bar{\mathbf{k}}_{\text{surf}}$ and $\bar{\mathbf{n}}$ are the unit vectors of the propagation direction and the outer normal of the medium, respectively, and we calculated the mean (integral) spin per one surface-mode particle. The momentum-dependent spin $\langle \mathbf{S}_{\text{surf}} \rangle$ originates from the transverse spin (5) of evanescent waves.

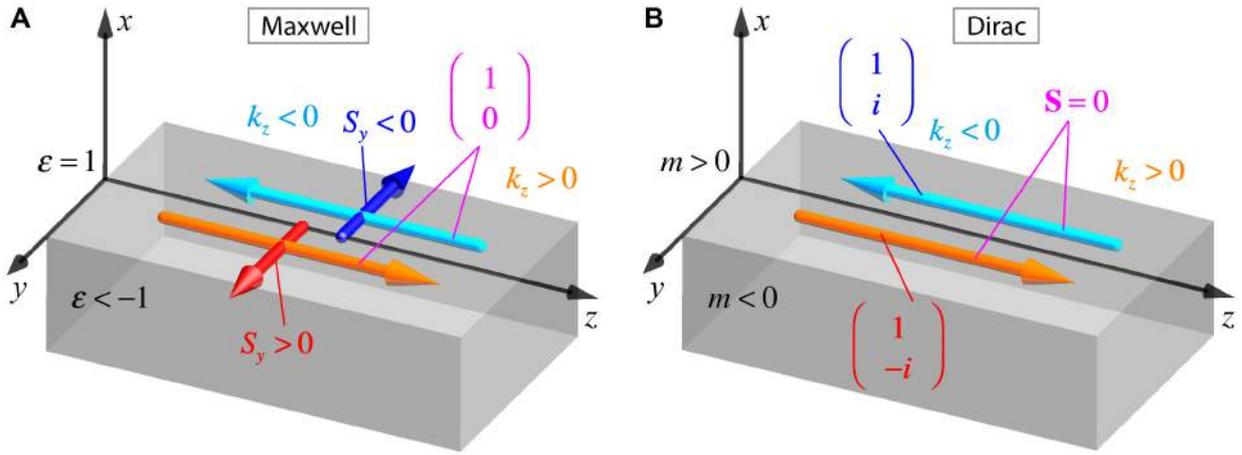

**Fig. 2. Spin and spinor properties of Maxwell and Dirac surface modes. (A)** Surface modes of Maxwell equations propagating along the interface between the vacuum and a non-transparent medium with $\mu = 1$, $\varepsilon < -1$. These surface waves have fixed polarization $\xi_{\text{surf}} = (1,0)^T$, but opposite transverse spins $\mathbf{S}$ locked to opposite wave momenta, Eqs. (5) and (6). **(B)** Topological surface modes of the Dirac equation at the interface between positive-mass and negative-mass regions [31,32]. These modes exhibit locking between their momenta and spinors: orthogonal polarizations propagate in opposite directions. However, the expectation value of their spin vanishes: $\mathbf{S}_{\text{surf}} = 0$ (see Supplementary Text).

Importantly, Eqs. (5) and (6) determine the momentum locking of the spin $\mathbf{S}$, but not of the polarization spinor $\xi$ (Fig. 2A). Polarization uniquely corresponds to spin for nonrelativistic electrons, but for relativistic particles these are different notions. The surface modes of Maxwell equations have momentum-dependent spin $\mathbf{S}_{\text{surf}}$ but fixed spinor $\xi_{\text{surf}}$, Eq. (6). The latter corresponds to the trivial $\mathbb{Z}_2$ invariant $\nu = 0$ and shows that surface Maxwell modes are bosons rather than helical fermions [30]. Nonetheless, these modes provide the unidirectional edge spin transport (QSHE) due to the spin $\mathbf{S}_{\text{surf}}$. Remarkably, precisely the opposite situation takes place in one of the main models for 3D electron topological insulators: the Dirac equation with surface modes at the interface between positive-mass and negative-mass regions (Fig. 2B) [31,32]. In this case, spinor-momentum locking occurs, which corresponds to the topological $\mathbb{Z}_2$ invariant $\nu = 1$. However, surprisingly, the expectation value of the spin of the surface Dirac modes vanishes due to the mutual cancellation of the polarization-dependent and polarization-independent (similar to Eq. (5)) contributions (see Supplementary Text). Thus, one can say that surface Maxwell modes exhibit unidirectional spin transport (QSHE) but with trivial $\mathbb{Z}_2$ spinor properties, while the surface Dirac modes are topologically-protected helical fermions which, however, do not transport spin. This is the second key point of our work.

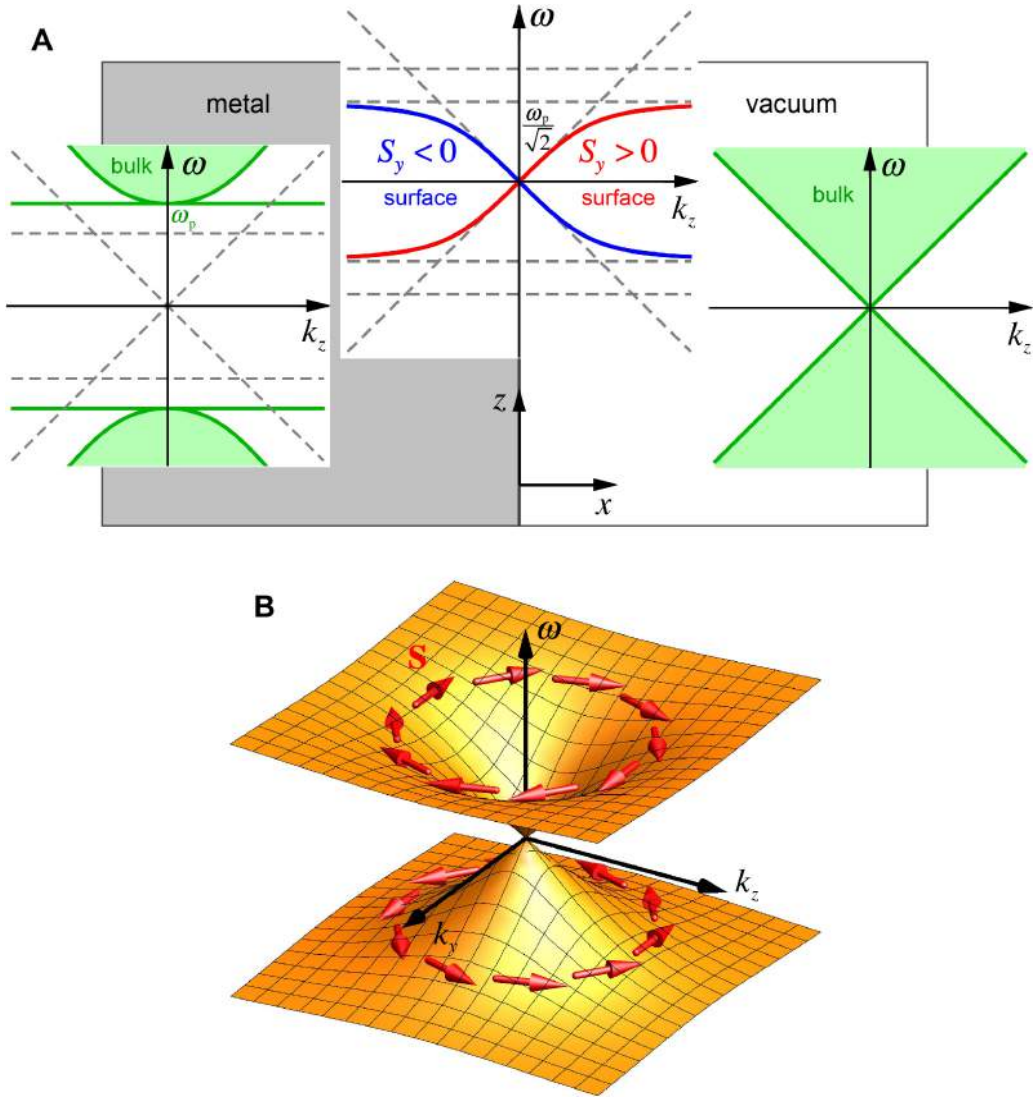

**Fig. 3. Dispersion and spin-momentum locking of surface plasmon-polaritons.** (**A**) Dispersion of bulk and surface modes at the vacuum-metal interface (see explanations in the text). SPPs exist inside the gap of the metal bulk spectrum and have spin-momentum locking associated with the transverse spin (5) and (6). (**B**) Two-dimensional dispersion of the same SPP mode exhibits a vortex spin texture similar to that for surface states of a 3D topological insulator [7,8].

Optical spin-momentum locking was recently observed in several experiments [22–27]. An important example is provided by surface plasmon-polaritons (SPPs) at the vacuum-metal interface [33]. Real metals are dispersive media with permittivity $\varepsilon(\omega) = 1 - \omega_p^2/\omega^2$, where $\omega_p$ is the plasma frequency. Metals are optical insulators at $\omega < \omega_p$, and at $\varepsilon < -1$ ($\omega < \omega_p/\sqrt{2}$) the vacuum-metal interface supports surface Maxwell modes, i.e., the SPPs, Fig. 3A. The metal becomes transparent at $\omega \geq \omega_p$, with bulk plasmons at $\omega = \omega_p$ and electromagnetic modes at $\omega > \omega_p$. Figure 3A shows that the vacuum-metal interface resembles, using condensed-matter analogies, the interface between a semimetal and an insulator. The SPP modes demonstrate spin-momentum locking (5) and (6) and non-removable (due to the light-cone spectrum in vacuum) spectral degeneracy at $k = 0$, which are typical for electron QSHE states. Furthermore, plotting the SPP spectrum for a 2D surface of a 3D metal (Fig. 3B), one can see the conical spectrum and vortex spin texture analogous to those in 3D electron topological insulators [7,8], but without the helical-fermion spinor properties (Fig. 2).

Figure 4 shows a schematic of the experiments [22–27] revealing spin-controlled unidirectional transport in electromagnetic surface or guided waves. A transversely-propagating free-space light beam with the usual spin $\mathbf{S}_{inc} = \sigma \bar{\mathbf{y}}$ (helicity $\sigma = \pm 1$) was coupled with the evanescent tails of the SPP or waveguide modes via some scatterer (e.g., a nanoparticle or an atom). In doing so, the opposite incident-spin states $\mathbf{S}_{inc} = \pm \bar{\mathbf{y}}$ excited the surface or guided modes running in the opposite directions: $\bar{\mathbf{k}}_{surf} = \pm \bar{\mathbf{z}}$. This spin-direction correlation reached almost 100% efficiency in various systems, independently of their details. This proves the universal spin-momentum locking in optical surface waves, i.e., the QSHE of light.

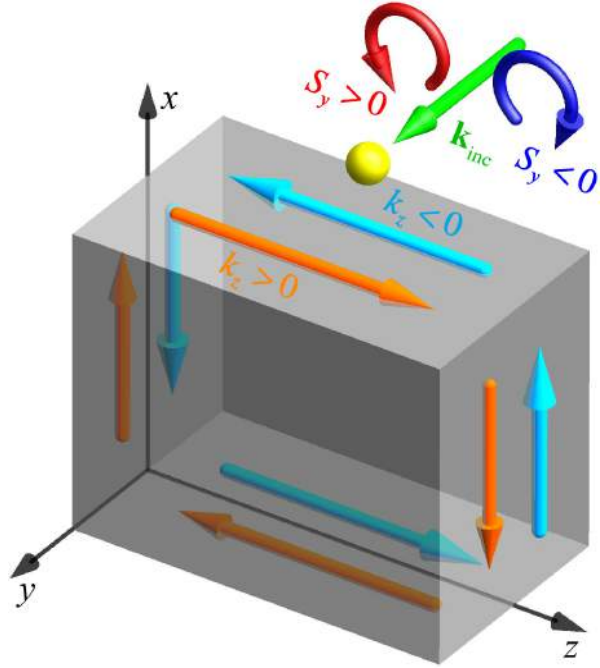

**Fig. 4. Schematic of experiments demonstrating the quantum spin Hall effect of light.** The incident $y$-propagating light (shown in green) is coupled to surface modes with evanescent free-space tails via some scatterer (e.g., a nanoparticle). Depending on the spin of the incident light, $\mathbf{S}_{inc} = \sigma \bar{\mathbf{y}}$ (the helicity $\sigma = \pm 1$ is shown here by the circular-polarization arrows), surface waves with opposite propagation directions $\bar{\mathbf{k}}_{surf} = \pm \bar{\mathbf{z}}$ are excited [22–27].

Thus, we have shown that light has intrinsic quantum spin Hall effect features, which originate from the spin-orbit interactions of photons. The corresponding spin-momentum locking originates solely from the basic properties of evanescent waves in Maxwell equations, and can be observed at any interface with the vacuum supporting surface or guided waves. In particular, surface plasmon-polaritons at a metal-vacuum interface exhibit features similar to those of surface states of topological insulators (i.e., vortex spin texture at the conical dispersion). Due to their trivial spinor structure, surface electromagnetic states are not helical fermions and are not protected from backscattering. Nonetheless, they do provide robust unidirectional spin transport. Our work shows that recent experiments, demonstrating highly efficient spin-controlled unidirectional excitation of surface or guided modes, can be interpreted as observations of the QSHE of light. Interestingly, the transverse spin locked to the direction of propagation seems to be a universal feature of surface vector waves of different nature. It appears in Maxwell and Dirac equations, as well as in Rayleigh surface waves in elastic media and surface-water waves. This offers robust angular-momentum-to-direction coupling in various surface waves as well as important analogies and generalizations involving quantum and classical wave theories.

**Acknowledgements:** We are grateful to Akira Furusaki, Yuri Bliokh, Elena Ostrovskaya, and Alexander Khanikaev for fruitful discussions. This work was partially supported by the RIKEN iTHES Project, MURI Center for Dynamic Magneto-Optics, JSPS-RFBR contract no. 12-02-92100 and a Grant-in-Aid for Scientific Research (S).

## Supplementary Text

**Evanescent modes and transverse spin in the Dirac equation**

One of the basic models of 3D topological insulators is based on the Dirac equation for a relativistic electron [31,32]. Therefore, we provide here an analysis of evanescent Dirac waves and their spin properties.

We write the 3D Dirac equation in the standard representation in units with $\hbar = c = 1$:

$$i\partial_t \psi = (\boldsymbol{\alpha} \cdot \hat{\mathbf{p}} + \beta m)\psi, \tag{S1}$$

where $\psi(\mathbf{r},t)$ is the Dirac bispinor, $\hat{\mathbf{p}} = -i\nabla_\mathbf{r}$ is the momentum operator, and $m$ is the mass. We also used the $4\times 4$ Dirac matrices:

$$\boldsymbol{\alpha} = \begin{pmatrix} \mathbf{0} & \boldsymbol{\sigma} \\ \boldsymbol{\sigma} & \mathbf{0} \end{pmatrix}, \quad \beta = \begin{pmatrix} \mathbf{I} & \mathbf{0} \\ \mathbf{0} & -\mathbf{I} \end{pmatrix},$$

where $\boldsymbol{\sigma}$ are the Pauli matrices

$$\sigma_x = \begin{pmatrix} 0 & 1 \\ 1 & 0 \end{pmatrix}, \quad \sigma_y = \begin{pmatrix} 0 & -i \\ i & 0 \end{pmatrix}, \quad \sigma_z = \begin{pmatrix} 1 & 0 \\ 0 & -1 \end{pmatrix},$$

and $\mathbf{I}$ and $\mathbf{0}$ are the $2\times 2$ identity and null matrices.

The plane-wave solutions of Eq. (S1) can be written as [34]

$$\psi_\mathbf{p}(\mathbf{r},t) = W(\mathbf{p})\exp[i(\mathbf{p}\cdot\mathbf{r} - Et)], \tag{S2}$$

$$W(\mathbf{p}) = \frac{1}{\sqrt{2}}\begin{pmatrix} \sqrt{1+\frac{m}{E}}\, w \\ \sqrt{1-\frac{m}{E}}\, \frac{\boldsymbol{\sigma}\cdot\mathbf{p}}{p}\, w \end{pmatrix}. \tag{S3}$$

Here $E = \pm\sqrt{p^2 + m^2}$ corresponds to two spin-degenerate energy bands (i.e., four bands in total), and $w = \begin{pmatrix} w_1 \\ w_2 \end{pmatrix}$, $w^\dagger w = 1$, is the normalized 2-component polarization spinor, which determines the spin state of the electron in its rest frame.

The spin density in Dirac waves can be obtained using the canonical spin operator $\boldsymbol{\Sigma} = \frac{1}{2}\begin{pmatrix} \boldsymbol{\sigma} & \mathbf{0} \\ \mathbf{0} & \boldsymbol{\sigma} \end{pmatrix}$. In particular, the spin density in the plane wave (S2) and (S3) is

$$\mathbf{S} = \psi^\dagger \boldsymbol{\Sigma} \psi = \frac{m}{E}\mathbf{s} + \left(1-\frac{m}{E}\right)\frac{\mathbf{p}(\mathbf{p}\cdot\mathbf{s})}{p^2}, \tag{S4}$$

where $\mathbf{s} = \frac{1}{2} w^\dagger \boldsymbol{\sigma} w$ characterizes the electron spin in the rest frame [34].

The evanescent-wave solution of Eq. (S1) can be written as a plane wave (S2) and (S3) with a complex momentum $\mathbf{p}$. Assuming an evanescent wave propagating in the $z$-direction and decaying in the positive $x$-direction, we have $\mathbf{p} = p_z \bar{\mathbf{z}} + i\kappa \bar{\mathbf{x}}$, where $p_z^2 - \kappa^2 = p^2 = E^2 - m^2$. Substituting this in Eqs. (S2) and (S3), we obtain

$$\psi_{\text{evan}} = W_{\text{evan}} \exp\left[ i(p_z z - Et) - \kappa x \right], \tag{S5}$$

$$W_{\text{evan}} = \frac{1}{\sqrt{2}} \begin{pmatrix} \sqrt{1+m/E}\, w_1 \\ \sqrt{1+m/E}\, w_2 \\ \sqrt{1-m/E}\, (p_z w_1 + i\kappa w_2)/p \\ \sqrt{1-m/E}\, (i\kappa w_1 - p_z w_2)/p \end{pmatrix}. \tag{S6}$$

We now calculate the spin density in the Dirac evanescent wave (S5) and (S6) akin to Eq. (S4). Representing the result in the general vector form yields

$$\mathbf{S} = \psi_{\text{evan}}^\dagger \boldsymbol{\Sigma} \psi_{\text{evan}} = \mathbf{S}_w + \mathbf{S}_\perp, \tag{S7}$$

$$\mathbf{S}_w = \left\{ \frac{m}{E} \mathbf{s} + \left(1 - \frac{m}{E}\right) \frac{\text{Re}\mathbf{p}(\text{Re}\mathbf{p}\cdot\mathbf{s})}{(\text{Re}\mathbf{p})^2} - \left(1 - \frac{m}{E}\right) \frac{(\text{Re}\mathbf{p} \times \text{Im}\mathbf{p})\left[(\text{Re}\mathbf{p} \times \text{Im}\mathbf{p})\cdot\mathbf{s}\right]}{p^2 (\text{Re}\mathbf{p})^2} \right\} e^{-2\text{Im}\mathbf{p}\cdot\mathbf{r}}, \tag{S8}$$

$$\mathbf{S}_\perp = \frac{1}{2}\left(1 - \frac{m}{E}\right) \frac{\text{Re}\mathbf{p} \times \text{Im}\mathbf{p}}{p^2} e^{-2\text{Im}\mathbf{p}\cdot\mathbf{r}}. \tag{S9}$$

This is a very interesting result, where we separated two distinct contributions $\mathbf{S}_w$ and $\mathbf{S}_\perp$. The first one, Eq. (S8), depends on the polarization spinor $w$ via the rest-frame spin vector $\mathbf{s}$, akin to the usual propagating-wave spin (S4). The second contribution, Eq. (S9), is a transverse spin, which is independent of the polarization and is entirely similar to the transverse spin in evanescent electromagnetic waves [20,21], Eq. (5). Note that, unlike the electromagnetic-wave case, the polarization-dependent spin $\mathbf{S}_w$ can also have a transverse component in the direction of $(\text{Re}\mathbf{p} \times \text{Im}\mathbf{p})$.

Thus, the above results show that the polarization-independent transverse spin $\mathbf{S}_\perp$ equally appears in evanescent waves in Maxwell and Dirac equations, Eqs. (5) and (S9).

**Jackiw-Rebbi surface modes of the Dirac equation**

A planar interface between two half-spaces, where 'electrons' are described by the Dirac equations with positive and negative masses, supports topological surface states. These modes are known as Jackiw-Rebbi solutions [31,32,35], see Fig. S1.

We consider the $x=0$ interface separating two regions with different electron masses:

$$m(x) = \begin{cases} m_1 > 0 \text{ for } x > 0 \\ m_2 < 0 \text{ for } x < 0 \end{cases}. \tag{S10}$$

The surface-state solutions have the form

$$\psi_{\text{surf}} = W_{\text{surf}} \exp\left[i(p_z z - Et) - \kappa_{1,2}|x|\right], \tag{S11}$$

where $p_z^2 - \kappa_{1,2}^2 = E^2 - m_{1,2}^2$ and the "1,2" subscripts correspond to the $x > 0$ and $x < 0$ half-spaces, respectively. Substituting Eq. (S11) into the Dirac equations (S1) with (S10), we find the decay factors $\kappa_1 = m_1$ and $\kappa_2 = -m_2$, as well as the dispersion relation $p_z = \pm E$, similar to the light cone. Hereafter, the "$\pm$" (or "$\mp$") signs indicate waves propagating in the positive and negative $z$-directions. The polarization bi-spinor of the mode (S11) can be written as

$$W_{\text{surf}} = \frac{1}{2\sqrt{N}} \begin{pmatrix} 1 \\ \mp i \\ \pm 1 \\ i \end{pmatrix}. \tag{S12}$$

Here $N = \dfrac{m_1 + |m_2|}{2m_1|m_2|}$ is chosen for the normalization $\int_{-\infty}^{\infty} \psi_{\text{surf}}^{\dagger} \psi_{\text{surf}} \, dx = 1$.

The bi-spinor (S12) corresponds to the polarization spinor $w_{\text{surf}} = \dfrac{1}{\sqrt{2}} \begin{pmatrix} 1 \\ \mp i \end{pmatrix}$, i.e., the rest-frame spin $\mathbf{s}_{\text{surf}} = \mp \dfrac{1}{2} \bar{\mathbf{y}}$. Thus, the rest-frame spin of the surface Jackiw-Rebbi mode is purely transverse, and opposite spins $s_y < 0$ and $s_y > 0$ are attached to the opposite directions of propagation $p_z > 0$ and $p_z < 0$. Moreover, the forward- and backward-propagating waves have orthogonal polarization spinors $w_{\text{surf}}$. This is intimately related to the fact that the topological Jackiw-Rebbi modes represent helical massless fermions, which have suppressed backscattering and are robust against disorder [7,8].

The above features are usually interpreted as spin-momentum locking in the quantum spin Hall effect and 3D topological insulators. However, quite surprisingly, the total spin density (S7)–(S9) vanishes identically in the Jackiw-Rebbi mode:

$$\mathbf{S}_{\text{surf}} = \psi_{\text{surf}}^{\dagger} \mathbf{\Sigma} \psi_{\text{surf}} \equiv 0. \tag{S13}$$

This happens because the transverse polarization-independent spin (S9) exactly cancels the polarization-dependent contribution (S8) for the above polarization:

$$\mathbf{S}_{\perp} = -\mathbf{S}_w = \pm \frac{1}{2N} \frac{E m_{1,2}}{(E + m_{1,2})^2} \bar{\mathbf{y}} e^{-2 m_{1,2} x}. \tag{S14}$$

Since both contributions above are locked to the momentum (as in the quantum spin Hall effect), one can say that opposite spin-momentum locking appears in the two spin contributions, while the total spin vanishes. Thus, the surface modes of the Dirac equation exhibit remarkable topological properties but do not provide any spin transport, i.e., the quantum spin Hall effect.

The cancellation of the spin in the Jackiw-Rebbi modes can be explained by the fact that the $m_1 = -m_2$ interface does not break spatial-inversion ($\mathcal{P}$) symmetry. Indeed, the $\mathcal{P}$ transformation results in the global change of the sign of the mass, which is not detectable. At the same time, the quantum spin Hall effect has broken inversion symmetry: it flips momenta but not spins. Thus, the quantum spin Hall effect is possible only in systems with broken inversion symmetry, such as the optical interfaces considered below.

For completeness, we determine the integral values of the two transverse spin contributions (S14). Denoting $\langle ... \rangle \equiv \int_{-\infty}^{\infty} ... \, dx$, we obtain

$$\langle \mathbf{S}_{\perp} \rangle = -\langle \mathbf{S}_w \rangle = \pm \frac{1}{2} \frac{E m_1 m_2 (2E + m_1 + m_2)}{(E + m_1)^2 (E + m_2)^2} \bar{\mathbf{y}}. \tag{S15}$$

**Surface waves in Maxwell equations**

Free-space Maxwell equations describe photons, i.e., massless spin-1 particles. These equations can also be written in the Dirac-like form with electric and magnetic fields forming an effective wavefunction [36]. For simplicity, in what follows we use natural electrodynamical units with $\varepsilon_0 = \mu_0 = c = 1$. The light-cone spectrum $\omega = k$ and bulk plane-wave solutions for Maxwell equations in vacuum are well-known. The complex electric-field amplitude of a plane wave is given by Eq. (1), whereas the corresponding magnetic field is $\mathbf{H} = \frac{\mathbf{k}}{k} \times \mathbf{E}$. The spin angular momentum of light is described by the following spin-1 operator [16,21,36]:

$$\hat{S}_x = -i \begin{pmatrix} 0 & 0 & 0 \\ 0 & 0 & 1 \\ 0 & -1 & 0 \end{pmatrix}, \quad \hat{S}_y = -i \begin{pmatrix} 0 & 0 & -1 \\ 0 & 0 & 0 \\ 1 & 0 & 0 \end{pmatrix}, \quad \hat{S}_z = -i \begin{pmatrix} 0 & 1 & 0 \\ -1 & 0 & 0 \\ 0 & 0 & 0 \end{pmatrix}. \quad (S16)$$

Using the operator $\hat{\mathbf{\Sigma}} = \begin{pmatrix} \hat{\mathbf{S}} & \mathbf{0} \\ \mathbf{0} & \hat{\mathbf{S}} \end{pmatrix}$ and the "wavefunction" $\psi = \frac{1}{2\sqrt{\omega}} \begin{pmatrix} \mathbf{E} \\ \mathbf{H} \end{pmatrix}$, normalized as $\psi^\dagger \psi = 1$, we obtain the spin density in a plane electromagnetic wave (1):

$$\mathbf{S} = \psi^\dagger \hat{\mathbf{\Sigma}} \psi = \sigma \frac{\mathbf{k}}{k}. \quad (S17)$$

In addition to the propagating "bulk" modes, Maxwell equations can also exhibit surface modes at interfaces between two homogeneous media with different electric permittivities $\varepsilon_{1,2}$ and permeabilities $\mu_{1,2}$. In the generic case, such an interface breaks the so-called dual symmetry between the electric and magnetic properties [29,37,38]. Therefore, instead of the circularly-polarized helicity eigenstates in free space, the eigenpolarizations of the surface modes are the transverse-electric (TE) and transverse-magnetic (TM) linearly-polarized waves. Considering waves propagating in the $(x,z)$-plane of a homogeneous medium, the full Maxwell equations can be reduced to uncoupled 2D scalar wave equations for the normal components $E_y$ (TE mode) and $H_y$ (TM mode):

$$\Delta E_y + \varepsilon \mu \omega^2 E_y = 0 \quad \text{(TE)},$$

$$\Delta H_y + \varepsilon \mu \omega^2 H_y = 0 \quad \text{(TM)}, \quad (S18)$$

where $\Delta = \frac{\partial^2}{\partial x^2} + \frac{\partial^2}{\partial z^2}$ and the other nonzero field components are determined by

$$H_z = -\frac{i}{\mu\omega} \frac{\partial E_y}{\partial x}, \quad H_x = \frac{i}{\mu\omega} \frac{\partial E_y}{\partial z} \quad \text{(TE)},$$

$$E_z = \frac{i}{\varepsilon\omega}\frac{\partial H_y}{\partial x}, \quad E_x = -\frac{i}{\varepsilon\omega}\frac{\partial H_y}{\partial z} \quad \text{(TM)}. \tag{S19}$$

Consider now the $x=0$ interface between two homogeneous media:

$$\varepsilon(x), \mu(x) = \begin{cases} \varepsilon_1, \mu_1 & \text{for } x > 0 \\ \varepsilon_2, \mu_2 & \text{for } x < 0 \end{cases}. \tag{S20}$$

We are looking for surface modes of the interface, which have the evanescent-wave form $\{E_y, H_y\} \propto \exp(ik_z z - \kappa_{1,2}|x|)$, where $\kappa_{1,2} = \left(k_z^2 - \varepsilon_{1,2}\mu_{1,2}\omega^2\right)^{1/2}$. Supplying the wave equations with the corresponding boundary conditions (continuity of the tangential components $E_{y,z}$ and $H_{y,z}$), one can derive the following equations for the TE and TM surface modes [39,40]:

$$\frac{\kappa_1}{\mu_1} + \frac{\kappa_2}{\mu_2} = 0 \quad \text{(TE)},$$

$$\frac{\kappa_1}{\varepsilon_1} + \frac{\kappa_2}{\varepsilon_2} = 0 \quad \text{(TM)}. \tag{S21}$$

Together with the dispersion relation $\kappa_{1,2} = \left(k_z^2 - \varepsilon_{1,2}\mu_{1,2}\omega^2\right)^{1/2}$, Eqs. (S21) determine the existence regions for the surface TE and TM modes [40]. These regions are depicted in Fig. S2 in the $(\varepsilon_2/\varepsilon_1, \mu_2/\mu_1)$-plane of the relative permittivity and permeability. Importantly, the TE and TM modes cannot exist simultaneously (except for the degenerate case $\varepsilon_2/\varepsilon_1 = \mu_2/\mu_1 = -1$). Therefore, instead of two polarization (spin) states of the propagating light, the surface modes of Maxwell equations have only one fixed polarization (either TE or TM) in the generic case.

The surface Maxwell modes have linear (conical) dispersion lying outside of the light cone for bulk waves. An example of such dispersion is show in Fig. S3. An interface between free space ($\varepsilon_1 = \mu_1 = 1$) and a medium with $\mu_2 = 1$ and $\varepsilon_2 < -1$ supports TM surface modes with the dispersion $k_z = \pm\sqrt{\frac{\varepsilon_1\varepsilon_2}{\varepsilon_1+\varepsilon_2}}\omega$ (hereafter we keep $\varepsilon_1$ for the sake of symmetry). Note that in this case, the medium 2 does not support any bulk modes and represents a perfect "insulator" for photons. Therefore, no spin properties (e.g., spin Chern number, etc.) can be ascribed to the $x < 0$ semispace. In addition, here we intentionally consider non-dispersive media with permittivities and permeabilities independent of $\omega$. Introducing a dispersion $\varepsilon_2(\omega)$, as in the case of a real metal, results in plasma-frequency longitudinal bulk modes (plasmons) and high-frequency transverse bulk modes (photons) in the medium 2. This is accompanied by the saturation of the surface-mode dispersion at lower frequencies, as shown in Fig. 3.

Both TE and TM surface modes are linearly-polarized, i.e., have zero helicity $\sigma = 0$ and no usual spin (S16). At the same time, they possess the transverse spin (5) [20,21]. In particular, for the above example of the vacuum-medium interface with $\mu_2 = 1$, $\varepsilon_2 < -1$, the spin of the TM surface wave is described by the canonical spin operator $\hat{\mathbf{\Sigma}}$ and the "wavefunction" in a medium $\psi = \dfrac{1}{2\sqrt{\omega}} \begin{pmatrix} \mathbf{E}/\sqrt{\mu} \\ \mathbf{H}/\sqrt{\varepsilon} \end{pmatrix}$ [21]. Since the electromagnetic energy density in the medium is $\dfrac{1}{4}\left(\varepsilon|\mathbf{E}|^2 + \mu|\mathbf{H}|^2\right)$, we impose the one-particle normalization $\int_{-\infty}^{\infty} \varepsilon\mu \left(\psi_{\text{surf}}^{\dagger} \psi_{\text{surf}}\right) dx = 1$ for the $x$-localized surface mode. This results in the following transverse spin density in the TM surface mode:

$$\mathbf{S}_{\text{surf}} = \psi_{\text{surf}}^{\dagger} \hat{\mathbf{\Sigma}} \psi_{\text{surf}} = \pm 2\omega \frac{-\varepsilon_{2,1}}{\left(\varepsilon_2^2 - \varepsilon_1^2\right)} \sqrt{\frac{\varepsilon_1 \varepsilon_2}{\varepsilon_1 + \varepsilon_2}} \, \overline{\mathbf{y}} \, e^{-2\kappa_{1,2}|x|}. \tag{S22}$$

This spin density is purely transverse, and the opposite spin directions $S_y > 0$ and $S_y < 0$ are locked with the opposite propagation directions $k_z > 0$ and $k_z < 0$. This is the main feature of the quantum spin Hall effect, which is emphasized in the main text of the paper. However, unlike the topological Jackiw-Rebbi modes, the surface electromagnetic modes have a fixed polarization which is independent of the momentum. In the above example, this is the TM-polarization described by the polarization spinor (Jones vector) $\xi = \begin{pmatrix} 1 \\ 0 \end{pmatrix}$. Thus, the surface modes of Maxwell equations are not helical fermions, they have usual scattering properties, and are not robust against disorder. Nonetheless, in contrast to the Jackiw-Rebbi modes, surface electromagnetic waves do provide robust unidirectional spin transport along the interface. Indeed, independently of the direction of propagation, the $z$-component of the spin current is determined by the product $S_y k_z$, which is always positive.

We emphasize that the transverse spin (S22) originates solely from the polarization-independent contribution $\mathbf{S}_\perp$, Eq. (5). Although the transverse-spin densities (S22) have opposite signs at the opposite sides of the interface, the total (i.e., $x$-integrated) transverse spin is non-zero:

$$\langle \mathbf{S}_{\text{surf}} \rangle = \pm \frac{1}{\sqrt{-\varepsilon_1 \varepsilon_2}} \, \overline{\mathbf{y}}. \tag{S23}$$

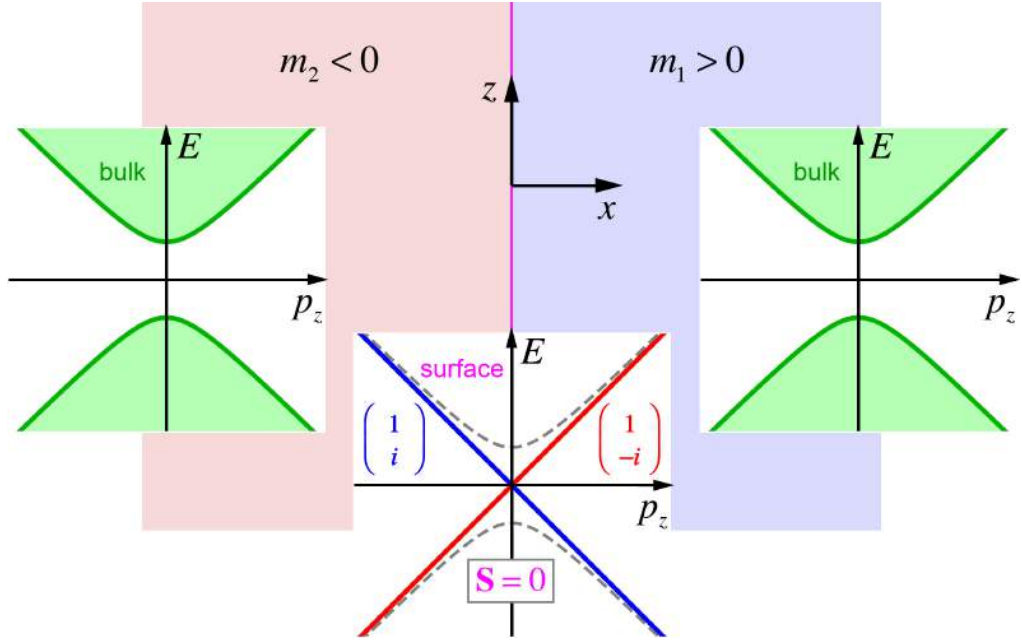

**Fig. S1.** Dispersions of the bulk and surface modes for the Dirac equation with the $x=0$ interface separating the half-spaces with positive and negative masses. In both half-spaces the Dirac bulk modes exist, while the interface supports topological Jackiw-Rebbi modes. These surface modes exhibit transverse spinor-momentum locking related to their helical fermion nature. However, the total transverse spin vanishes identically due to the mutual cancellation of the polarization-dependent and polarization-independent contributions, Eqs. (S7)–(S9) and (S13)–(S15).

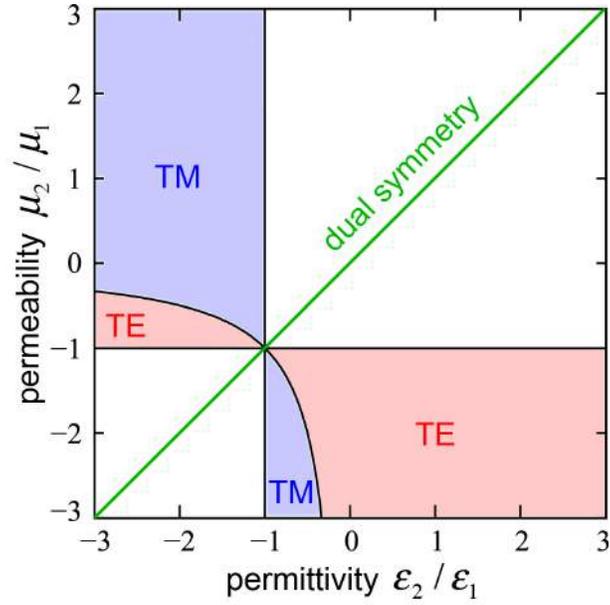

**Fig. S2.** The existence regions for surface Maxwell modes at the planar interface between two media with permittivities $\varepsilon_{1,2}$ and permeabilities $\mu_{1,2}$. Either transverse-electric (TE), or transverse-magnetic (TM) linearly-polarized modes, or none of these can exist at every value of the parameters. (The only exclusion is the degenerate case $\varepsilon_2/\varepsilon_1 = \mu_2/\mu_1 = -1$.) Moreover, the dual symmetry between electric and magnetic properties (shown by the green line) must be broken for the existence of surface modes of Maxwell equations.

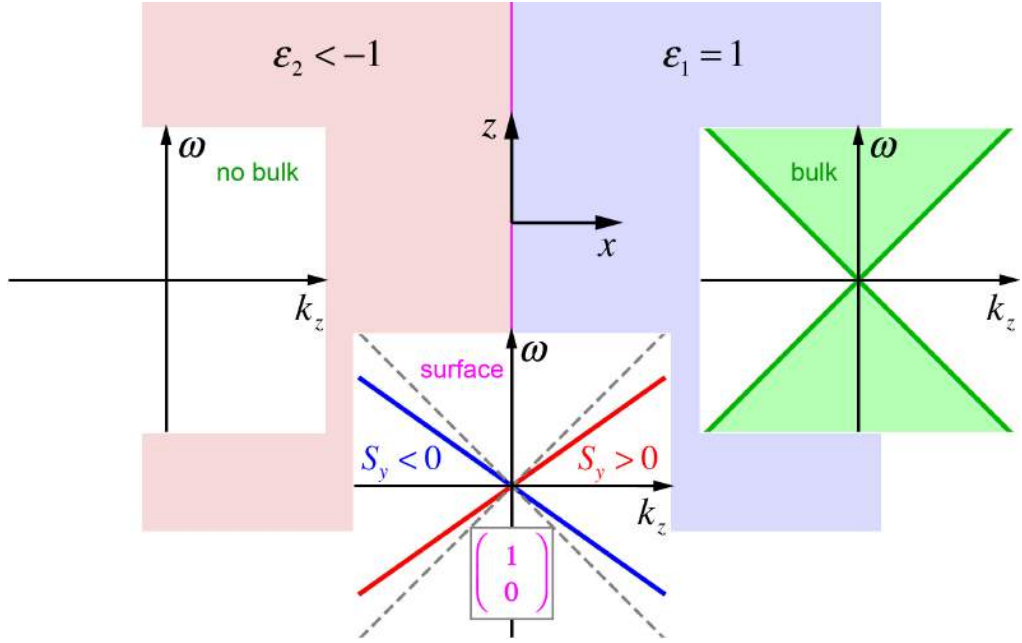

**Fig. S3.** Dispersions of the bulk and surface modes of Maxwell equations with the $x=0$ interface separating two half-spaces with different permittivities $\varepsilon$ and the same permeability $\mu=1$. While the usual light represents bulk modes of the vacuum, there are no propagating waves in the medium with negative permittivity. The interface supports surface modes with the fixed TM polarization described by the Jones-vector spinor $\xi = \begin{pmatrix} 1 \\ 0 \end{pmatrix}$. These modes exhibit transverse spin-momentum locking due to the transverse polarization-independent spin, Eqs. (5), (6), (S22) and (S23). This corresponds to the quantum spin Hall effect of light. Furthermore the spectrum of these modes lies outside the light cone and therefore has a non-removable degeneracy at $k=0$.